\documentstyle[aps,12pt,amssymb]{revtex}
\newenvironment{proof}{{\bf Proof:}}{}
\newtheorem{theorem}{Theorem}[section]
\newtheorem{lemma}[theorem]{Lemma}

\newtheorem{corollary}[theorem]{Corollary}
\newtheorem{remark}{Remark}[section]
\newtheorem{definition}{Definition}[section]
\newcommand{\qed}{$\bf \Box$}

\newcommand{\eps}{\varepsilon}

\def\C{{\bf C}}

\begin{document}
\title{Integrating Singular Functions on the Sphere}
\author{Alicia Dickenstein$^{\small 1}$
\thanks{ Supported by UBACYT and CONICET.},
Mirta Susana Iriondo$^{\small 2}$  
\thanks{Supported by STINT. } and \\
Teresita Alejandra Rojas$^{\small 2}$
\thanks{Fellow of CONICOR.}\\
{\small${}^{\small 1}$ Dto. de Matem\'atica, FCE y N, UBA,}\\
{\small Ciudad Universitaria, Pab. I , 1428 Buenos Aires, Argentina}\\
{\small ${}^{\small 2}$ FaMAF, Medina Allende y Haya de la Torre,}\\
 {\small Ciudad Universitaria, 5000 C\'ordoba, Argentina}}
\date{}
\maketitle

\begin{center}
Abstract
\end{center}

\begin{abstract}
We obtain rigorous results concerning 
the evaluation of integrals on the two sphere using complex methods. 
It is shown that for regular as well as singular
functions  which admit poles, 
the integral can be reduced to the calculation of residues through a 
limiting procedure.
\end{abstract}

\section{Introduction}
\label{sec1}
Standard textbooks on mathematical physics state that integrals of regular 
functions on the sphere are easy to compute if one uses  the spherical 
harmonics decomposition. These books, however, do not explain what to do if 
the integrands have singularities since in that case the functions do not 
admit an expansion in spherical harmonics. 

In this paper we formulate a useful technique to evaluate integrals on the 
two sphere of integrands that might possess singular (pole-type) behavior. 
The method is based on the use of Stokes Theorem to convert the two-dimensional
integrals on the complex plane (obtained by the stereographic projection of the
sphere), into  line integrals
around singular points and then to evaluate  the latter by a
generalization of the Cauchy Residue Theorem. Our approach is inspired in
the theory of Residue Currents developed in \cite{herrera}.

Our main result (stated in Theorem \ref{main-theorem}) is that 
principal values of integrals in the sphere of (not necessarily integrable)
$C^n$ functions $g$ with
poles (cf. Definition \ref{main-def}) can be explicitly evaluated as a sum 
of residue limits of a $C^n$
solution  with poles of the differential equation
${\partial\over\partial \bar{z}}f=\frac{g}{(1+z\bar z)^2}.$
We also show how to construct an explicit $C^n$ solution with poles to this 
equation (cf. Corollary \ref{main-cor}). These results are useful in a wide 
range of 
different physical theories. One can apply these techniques to obtain 
explicit evaluations of Feynman propagators and Feynman graphs,  to obtain 
solutions of differentials equation on the sphere, etc.

This paper is organized as follows: In section \ref{sec2} we give some 
mathematical preliminaries needed for the present work. In section \ref{sec3} 
we present the main result for functions that are $C^n$ on $\C$ except for a 
finite number of singularities, and in the last section we apply these results 
to integrals on the sphere.

\section{Preliminaries}
\label{sec2}

For later reference we   give some standard  formulae, and  review  
some results which contain the basic ideas that we shall use throughout  
this paper (for more details see \cite{lang}).

Let  $D$ be a closed disc in the complex plane $\C$, bounded by the circle
$\gamma$. The Cauchy formula
\begin{equation}
h(z_0)=\frac{1}{2\pi i}\oint_{\gamma}\frac{h(z)\mbox{d}z}{z-z_0},
\label{eq:cauchy1}
\end{equation}
gives the value of $h(z_0)$ for any $z_0$ in the interior of $ D$ 
as an integral along  $\gamma$ with 
counterclockwise orientation when $h$ is a holomorphic (complex differentiable)
function 
on some open neighborhood of $D$. But if $h$ is not 
holomorphic but merely smooth  (i.e. its real and imaginary parts are 
continuously differentiable in the real sense) there is  a similar formula 
giving the value of $h(z_0)$, which  shall be given later in this section.

Let us write $z=x+iy$, and let
$$h(z,\bar z)=h_1(x,y)+ih_2(x,y),$$
where $h_1$ and $h_2$ are the real and imaginary parts of $h$ 
respectively.  
We say that $h$ is $C^\infty$ ($C^n$) if $h_1$ and $h_2$ are   
$C^\infty$ ($C^n$) in the usual  sense for functions of two real 
variables $x$ and $y$.  In other words,  all partial derivatives of any 
order (all those up to  the $n$-th order) of $h_1$ and $h_2$,  
exist and are continuous. We write $h\in C^n(D) $ to mean that $h$ is $C^n$ 
on some open set containing $D$.

For such functions we define 
$$
{\partial h\over\partial z}={1\over2}\bigg ({\partial h\over\partial x}-i
{\partial h\over\partial y}\bigg) \quad\mbox{and}\quad
{\partial h\over\partial \bar{z}}={1\over2}\bigg ({\partial h\over\partial x}+i
{\partial h\over\partial y}\bigg).
$$

Thus, the {\bf Cauchy-Riemann equations} can be formulated by saying that 
$h$ is holomorphic if and only if 
\begin{equation}
{\partial h\over\partial\bar{z}}=0. 
\label{holo}
\end{equation}

Since  $z=x+iy$ and $\bar z=x-iy$, a $C^\infty$ or 
$C^n$complex function  can be described as a function in the complex variables $(z, \bar z)$, 
for which  the  following complex formulation of the Green-Stokes formula
holds.

\vspace{.2in}
\noindent {\bf Green-Stokes Formula}. Given a region $B\subset \C$ bounded by 
a finite number of curves,
oriented so that the region lies to the left of each curve, the  
Stokes-Green Formula  in the variable $z$ is
\begin{equation}
\int_{\gamma} f dz+g \mbox{d}\bar{z}= \int_B \bigg ({\partial g\over \partial
z} -{\partial f \over \partial \bar{z}} \bigg )\mbox{d}z\wedge \mbox{d}\bar{z},
\label{Stokes1}
\end{equation}
where $\gamma$ is the boundary of $B$,  $f$ and $g$ have continuous first partial derivatives,
$$
\mbox{d}z=\mbox{d}x+i\mbox{d}y\qquad \mbox{d}\bar{z}=\mbox{d}x-i\mbox{d}y
$$
and
\begin{equation}
\mbox{d}z\wedge \mbox{d}\bar{z}=-2i\mbox{d}x\wedge \mbox{d}y.
\label{volume}
\end{equation}

\vspace{.2in}
Using  (\ref{Stokes1}), one can formulate  the following version of the Cauchy 
Theorem for $C^1(D)$ 
functions (see \cite{lang} and \cite{dolbeault} for a proof).

\vspace{.2in}
\noindent{\bf Cauchy Theorem for $C^1(D)$ functions}.
{\it Let $h\in C^1(D)$ and  $z_0$ be a point in the interior of $D$. 
Further, let 
$\gamma$ be the circle around $D$ with counterclockwise orientation, then }
\begin{equation}
h(z_0,\bar z_0)={1\over 2\pi i}\oint_{\gamma} {h\over z-z_0}
\mbox{d}z+{1\over 2\pi i}\int_D  {\partial h\over\partial\bar{z}}{\mbox{d}z
\wedge \mbox{d} \bar{z}\over z-z_0}.
\label{lang}
\end{equation}

\vspace{.2in}
It is clear that if $h$ is holomorphic, the double integral disappears  
obtaining the standard Cauchy formula.

In the proof of this Theorem the basic ingredient is the evaluation of  the improper  
integral 
\begin{equation}
\int_D  {\partial h\over\partial\bar{z}}{\mbox{d}z\wedge \mbox{d}
\bar{z}\over z-z_0}=\lim_{\varepsilon\to 0}\int_{D^\varepsilon} {\partial 
h\over\partial\bar{z}}{\mbox{d}z\wedge \mbox{d}
\bar{z}\over z-z_0},
\label{eq:inteps}
\end{equation}
where $D^\varepsilon$ is the region obtained from $D$ by deleting a small disc 
of radius $\varepsilon$ centered at the point $z_0$. The boundary of this region,
$\partial D^\varepsilon$,  consists of two curves, $\gamma$ and $-\gamma_\varepsilon$,
where the first one  has counterclockwise orientation and the second one has 
clockwise orientation.


\begin{remark} Note that since $h$ is $C^1$ the l.h.s. of (\ref{eq:inteps}) 
exists.
\end{remark}
Therefore,  using the Stokes formula (\ref{Stokes1}),  we obtain 
$$
\int_{D^\varepsilon} {\partial 
h\over\partial\bar{z}}{\mbox{d}z\wedge \mbox{d}
\bar{z}\over z-z_0}=\oint_{\gamma_\varepsilon} {h\over z-z_0}
\mbox{d}z-\oint_{\gamma} {h\over z-z_0}
\mbox{d}z.
$$
The integral along $\gamma_\varepsilon$ is evaluated replacing $h(z,\bar z)$ by its 
$0$-order  Taylor expansion in the variables $(z,\bar z)$, which exists  
since $h$ is  $C^1$, i.e.
\begin{equation}
\label{eq:taylor}
\oint_{\gamma_\varepsilon}\frac{h(z,\bar z)}{z-z_0}=\oint_{\gamma_\varepsilon}
\frac{h(z_0,\bar z_0)}{z-z_0}+\oint_{\gamma_\varepsilon}\sum_{m+l=1}\Gamma_{ml}(z,\bar z)\;
(\bar{z}-\bar{z}_0)^m(z-z_0)^{l},
\end{equation}
for some continuous functions $\Gamma_{10}$ and $\Gamma_{01}.$
Taking the limit of (\ref{eq:taylor}) when
$\varepsilon \to 0$,  the second integral
of the r.h.s. of (\ref{eq:taylor})   is proved to 
be zero, and   the first one  gives us  the l.h.s. of 
 (\ref{lang}).

Considering the special case when the function $h$ vanishes on the boundary of 
the disc, the integral along the circle $\gamma$ is equal to 0, thus the 
formula (\ref{lang}) becomes
$$
h(z_0,\bar z_0)={1\over 2\pi i}\int_D  {\partial h\over\partial\bar{z}}{\mbox{d}z
\wedge \mbox{d}
\bar{z}\over z-z_0}.
$$
This allows us to recover the values of a  function with compact  
support from its derivative 
$\displaystyle{\frac{\partial h}{\partial\bar{z}}}$. 
Conversely, one gets the following result (c.f. \cite{dolbeault}).

\begin{theorem}
\label{green} 
Let $h\in C^1(D)$ be a $C^1$ function on the closed disc $D$, 
with compact support contained in the interior of $D$. Then the function 
$$
f(z,\bar z)={1\over 2\pi i}\int_D \frac {h(\zeta,\bar\zeta)}{\zeta-z}
\mbox{d}\zeta\wedge \mbox{d}\bar \zeta
$$
is defined and is  $C^1$ on $D$, and satisfies
$$
\frac{\partial f}{\partial\bar{z}}= h(z,\bar z)\qquad\mbox{for }\qquad z\in D.
$$
\end{theorem}

The proof is essentially a Corollary of the Cauchy Theorem by means of 
differentiation under the integral sign after the change of variable 
$\eta=\zeta-z$. 

\begin{remark}
In case $h$ is a function with compact support contained in the interior of 
$D$, which has continuous partial derivatives up to some order
$n \geq 1$, $f$ is also a $C^n$ function on $D$. 
\label{ene}
\end{remark}

\section{Some results in Complex Analysis}
\label{sec3}

In this section we shall extend the Cauchy Theorem and Theorem 
\ref{green} to functions that  are $C^n$ in  an open set obtained  by 
removing a finite number of points in ${\bf C}$, with singularities
of the following kind:

\begin{definition}
Let $n\in {\bf N}$ and ${\cal F}$ be a finite set of points in $\C$, 
i.e. ${\cal F}=\big \{z_i\big \}_{i=0}^s$. We say that a function 
$f \in C^n({\bf C}-{\cal F})$, is a {\it $C^n$ function with poles} 
if:
\begin{enumerate}
\item{} For each $z_i$ there exists
a neighborhood ${\cal N}_i$ of $z_i$ and a $C^n$ function  $h$
on ${\cal N}_i $ such that  
$f(z,\bar z) = {h(z,\bar z)\over(z-z_i)^m}$,  for some integer $m, \, m\leq n$.
\item{} There exists a positive number $\eps$ and a
$C^n$ function  $h$ on the ball of radius $\eps$ around the origin
such that for any $z, \ 0 < |z| < \eps ,$  
$f(1/z,1/\bar z) = {h(z,\bar z)\over z^m}$,  for some integer $m,\, m\leq n-2$.
\end{enumerate}
\label{def:pole}
\end{definition}

\begin{remark}
The above definition is in fact concerned with the singular
behavior of the $1$-form $f(z,\bar z) dz$ on $\C \cup \{\infty\}$,
which may be identified with the $2$-sphere $S^2.$
The subtle point on the behavior at infinity is that the function
$h$, which is in principle defined in a punctured ball around the
origin, has in fact a $C^n$ extension to the entire ball.
\end{remark}

We now present the tools needed to generalize 
the Cauchy Theorem for $C^n$ functions with poles.
This generalization is stated in Theorem \ref{cauchy2}  and its proof
is based on the same basic idea as the proof of the Cauchy Theorem for 
$C^1({D})$ functions given in the previous section.
Let $f$ be a $C^n$ function with poles such 
that $\displaystyle{\partial f\over \partial \bar{z} }$ 
is an integrable function on $\C$. We can then write
\begin{equation}
\label{eq:PV}
\int_{\bf C}  {\partial f\over\partial\bar{z}}\mbox{d}z\wedge \mbox{d}\bar z
=\lim_{{R\to \infty} \atop {\varepsilon\to 0}}\int_{D^\varepsilon_R} {\partial f\over
\partial \bar{z}}\mbox{d}z\wedge \mbox{d}
\bar{z},
\end{equation}
where $D^\varepsilon_R$ is obtained by first taking  from the disc $D_R$, 
centered at 
the origin with radius $R$ sufficiently large so that the poles of $f$ 
remain inside $D_R$, and then deleting discs of radius $\varepsilon$ around 
the poles.

\begin{remark}
Even if $\displaystyle{\partial f\over \partial \bar{z} }$ is not an 
integrable function over $\C$, the limit in the r.h.s of (\ref{eq:PV})  
does anyway exist (cf. \cite{herrera}). 
\end{remark}
This Remark justifies the following Definition:

\begin{definition}
Let $f \in C^n({\bf C}-{\cal F})$. 
We define the principal value (PV) of the integral of $f$ over ${\bf C}$ as
$$
{\rm PV}
\int_{\C} f(z,\bar z)\mbox{d}z\wedge \mbox{d}\bar{z}:=
\lim_{{R\to \infty} \atop {\varepsilon\to 0}}\int_{D^\varepsilon_R}   
f(z,\bar z)\mbox{d}z\wedge \mbox{d}\bar{z}.
$$
\end{definition}
Hence, in order to obtain a formula similar to (\ref{lang}), we  shall 
evaluate the integral
$$
{\rm PV}
\int_{\C} {\partial\over \partial \bar{z} }f(z,\bar z)\mbox{d}z\wedge \mbox{d}\bar{z}=
\lim_{{R\to \infty} \atop {\varepsilon\to 0}}\int_{D^\varepsilon_R}   
{\partial\over \partial \bar{z} }f(z,\bar z)\mbox{d}z\wedge \mbox{d}\bar{z}.
$$
using the Stokes-Green formula (\ref{Stokes1}) over $D_R^\varepsilon$: 
$$
\int_{D^\varepsilon_R}{\partial \over\partial\bar{z}} f(z,\bar z)\mbox{d}z\wedge 
\mbox{d}\bar{z}=-\int_{\partial D^\varepsilon_R} f(z,\bar z)\mbox{d}z,
$$
where $\partial D_R^\varepsilon$ is the oriented boundary of $D^\varepsilon_R$ 
given by 
$$
\partial D^\varepsilon_R=\cup_l(-\gamma_\varepsilon^l)\cup\gamma_R
$$
and $\gamma_\varepsilon^l$ and $\gamma_R$ are circles, the first ones centered 
at the poles $z=z_l$ with radius $\varepsilon$ and the second one being the 
boundary of $D_R$, all with counterclockwise orientation. Therefore,
\begin{equation}
\label{eq:Stokes2}
\int_{D^\varepsilon_R}{\partial \over\partial\bar{z}} f(z,\bar z)\mbox{d}z\wedge 
\mbox{d}\bar{z}=\sum_l \oint_{\gamma_\varepsilon^l} f(z,\bar z)\mbox{d}z-\oint_{\gamma_R}
f(z,\bar z)\mbox{d}z.
\end{equation}

The integrals along $\gamma_\varepsilon^l$ can be evaluated in the following way.
By Definition \ref{def:pole},  a  function with a pole  of multiplicity $n$ at  
$z_0$ can be written as 
\begin{equation}
\label{eq:npole}
f(z,\bar z)= {h(z,\bar z)\over(z-z_0)^n},
\end{equation}
with $h(z,\bar z)\in C^n(D)$ for some $D$ containing $z_0$.  

As $h(z,\bar z)\in C^n({D})$, it has  a Taylor expansion  up to the $(n-1)$-th order. 
The integral containing the remainder along  $\gamma^l_\varepsilon$ becomes zero 
at the limit when $\varepsilon \to 0$ and the integral containing the Taylor 
polynomial  gives rise to the definition of the residue of $C^n$  functions 
with poles.

The above assertion is proved in the following Lemma and  its proof  is  
standard (c.f. \cite{alicia}).

\begin{lemma}
Let $f$ be a $C^n$ function with poles and write $f$ as in (\ref{eq:npole})
near a pole $z_0.$ Then
$$
\lim_{\varepsilon\to 0}\oint_{|z-z_0 |=\varepsilon}f(z,\bar z)\mbox{d}z=
{2\pi i\over(n-1)!}\partial_z^{(n-1)} h(z_0,\bar z_0),
$$
\label{lemma2}
\end{lemma}

\begin{proof}
Since $h$ is a $C^n$ function in a neighborhood of $z_0$, we may consider
its Taylor expansion 
$$
h(z,\bar z)=\sum_{0\leq m+l<n}{\partial_{\bar{z}}^m\partial_{z}^l\over m!\; l!}
h(z_0,\bar z_0)\; (\bar{z}-\bar{z}_0)^m(z-z_0)^l+\sum_{m+l=n} 
\Gamma_{lm}\; (\bar{z}-\bar{z}_0)^m(z-z_0)^l,
$$
where $\Gamma_{ml}$ are continuous functions in $(z,\bar z)$.
 It is obvious that
$$
\lim_{\varepsilon\to 0}\oint_{|z-z_0 |=\varepsilon} \Gamma_{lm}(z,\bar z)\; 
(\bar{z}-\bar{z_0})^m(z-z_0)^{ l-n}\mbox{d}z=0
$$
since $\Gamma_{lm}(z,\bar z)\; (\bar{z}-\bar{z}_0)^m(z-z_0)^{ l-n}$ is bounded 
near $z_0$ and the length of the path goes to zero when $\varepsilon \to 0$. 
Thus 
\begin{equation}
\label{eq:limeps}
\, \lim_{\varepsilon\to 0}\oint_{|z-z_0 |=\varepsilon}f(z,\bar z)\mbox{d}z=
\sum_{0\leq m+l<n}{\partial_{\bar{z}}^m\partial_{z}^l\over m!\; l!}h(z_0,\bar z_0) 
\;\lim_{\varepsilon\to 0}\oint_{|z-z_0 |=\varepsilon} (\bar{z}-\bar{z}_0)^m
(z-z_0)^{l-n}\mbox{d}z.
\end{equation}
In case  $m> 0$, we claim that all the line integrals are zero. In order to 
see this, consider the integral 
$$
I_\varepsilon=\oint_{|z-z_0 |=\varepsilon} (\bar{z}-\bar{z}_0)^m
(z-z_0)^{l-n}\mbox{d}z.
$$
In polar coordinates  this integral can be written as 
$$
I_\varepsilon=i\int_0^{2\pi}\varepsilon^{l-n+m+1}\mbox{e}^{i(l-n-m+1)t} \mbox{d}t.
$$
It is clear that $ I_\varepsilon=0$ if $l-n-m+1\neq 0$; otherwise $l-n+
m+1=2m$ which implies that $\lim_{\varepsilon\to 0}I_\varepsilon=0$. 
Thus,  the only nonvanishing terms of (\ref{eq:limeps}) are those given by 
setting $m=0$.

In this case the only non zero integral is when $l-n=-1$.  Therefore
$$
\lim_{\varepsilon\to 0}\oint_{|z-z_0 |=\varepsilon}f(z,\bar z)\mbox{d}z={2\pi 
i\over(n-1)!}\partial_z^{(n-1)} h(z_0,\bar z_0).
$$
This completes the proof. \qed
\end{proof}

\vspace*{.2in}
This result gives rise to the definition of  local residue for  $C^n$ functions
with poles.

\begin{definition}
Let $f$ be a function with a pole at $z_0$.
We define the {\it residue of $f$ at $z_0$} by the limit
$$
{\rm Res}\bigg(f;z_0\bigg):= {1 \over 2\pi i}
\lim_{\varepsilon\to 0}\oint_{|z-z_0|=\varepsilon}f(z,\bar z)\mbox{d}z.
$$
\end{definition}
Note that when the function  has a holomophic numerator, this definition is 
the standard one for meromophic functions.

Moreover, we define:

\begin{definition}
Given a $C^n$ function with poles, the {\it residue of $f$ at infinity} is 
the limit
$$
{\rm Res}\bigg(f;\infty\bigg ):= -{1 \over 2\pi i}
\lim_{R\to \infty}\oint_{\gamma_R}f(z,\bar z)\mbox{d}z.
$$
\end{definition}

\medskip

\begin{remark}
Given a $C^n$ function with poles $f$, similar arguments as in the
proof of Lemma \ref{lemma2} after the change of variable given
by the inversion $1/z$, show that the above limit exists, i.e.
the residue of $f$ at infinity is well defined.
\label{poleinf} 
\end{remark}

\vspace*{.2in}
\noindent We now generalize the Cauchy Theorem to $C^n$ functions with poles.
 
\begin{theorem}
Let $f$ be a $C^{n}$ function with poles, 
regular on $\C-{\cal F}$, where 
${\cal F}=\{
z_1,\ldots,z_s\}$. Then, the {\it principal value} 
$$
{\rm PV}  
\int_{\C} {\partial \over\partial\bar{z}}
 f(z,\bar z)\mbox{d}z\wedge \mbox{d}\bar{z}
$$
exists and  
$$
{\rm PV}
\int_{\C}{\partial \over\partial\bar{z}}
 f(z,\bar z)\mbox{d}z\wedge \mbox{d}\bar{z}=2\pi i\bigg (\sum_l {\rm Res}( f;z_l)
+{\rm Res} (f;\infty)\bigg ). 
$$
\label{cauchy2}
\end{theorem}

\begin{proof}
By equation (\ref{eq:Stokes2})
$$
{\rm PV}\int_{\C}{\partial \over\partial\bar{z}} f(z,\bar z)\mbox{d}z\wedge 
\mbox{d}\bar{z}=\sum_l \lim_{\varepsilon\to 0}\oint_{\gamma_\varepsilon^l} 
f(z,\bar z)\mbox{d}z-\lim_{R\to\infty}\oint_{\gamma_R}f(z,\bar z)\mbox{d}z,
$$
and the limits in the r.h.s. exist by Lemma \ref{lemma2} and
Remark \ref{poleinf}, and equal $2 \pi i$ times the corresponding
residues of $f$.
The statement follows at once.\qed
\end{proof}

\vspace*{.2in}
Considering the special case when $f$ has a simple pole at $z_0$, i.e.  
$$
f(z,\bar z)={h(z,\bar z)\over (z-z_0)},\qquad \mbox{with}\qquad h\in C^n({D})
$$
and Res$(f,\infty)=0$, Theorem \ref{cauchy2}  gives
$$
h(z_0,\bar z_0)={1\over 2\pi i}{\rm PV}\int_{\C}{\partial \over\partial\bar{z}}h
(z,\bar z){\mbox{d}z\wedge \mbox{d}\bar{z}\over(z-z_0)}.
$$
This allows to recover the values of the function $h$ from its derivatives.
Conversely, as in the case of $C^1(D)$ functions with compact support in $D$, 
one has the following result.

\begin{theorem}
Let $g$ be a $C^1(\bf C)$  function satisfying
\begin{equation}
\lim_{|z|\to \infty} |z|^{1+\alpha} |g(z,\bar z)| =0,
\label{decay}
\end{equation}
for some positive $\alpha.$ Then the function 
$$
f(z,\bar z)={1\over 2\pi i}\int_\C g(\zeta,\bar \zeta){\mbox{d}\zeta\wedge 
\mbox{d}\bar{\zeta}\over (\zeta-z)}
$$
is in $C^1(\bf C)$ and satisfies 
$\displaystyle{{\partial \over\partial\bar{z}}f =g}$ for $ z\in \C$.
\label{intrep}
\end{theorem}

\begin{proof}
Since ${1 \over \zeta - z}$ is integrable near $z$ and tends to $0$ for 
$|\zeta| \to \infty$ and $g$ has the decay given by  (\ref{decay}), we deduce
the existence of the integral. With the change of variables  $\eta=\zeta-z $ 
the integral becomes
$$
f(z,\bar z)={1\over 2\pi i}\int_\C g(z+\eta,\bar z+\bar \eta)
{\mbox{d}\eta\wedge \mbox{d}\bar{\eta}\over  \eta}
$$
and applying  $\displaystyle{{\partial \over\partial\bar{z}}}$ to $f$, 
we obtain
\begin{eqnarray*}
{\partial \over\partial\bar{z}}f(z,\bar z)&=&{1\over 2\pi i}\int_\C {\partial 
\over
\partial\bar{z}}g (z+\eta,\bar z+\bar \eta)
{\mbox{d}\eta\wedge \mbox{d}\bar{\eta}\over \eta}\\
&=&{1\over 2\pi i}\int_\C {\partial \over 
\partial\bar{\zeta}}
 g(\zeta,\bar \zeta)
{\mbox{d}\zeta\wedge \mbox{d}\bar{\zeta}\over  (\zeta-z)}.
\end{eqnarray*}

Because of the decay of $g$, the residue of
${g(\zeta,\bar \zeta) \over (\zeta - z)}$ at infinity is zero. Thus,  
applying Theorem \ref{cauchy2}, we obtain 
$\displaystyle{{\partial \over\partial\bar{z}}f=g}$.\qed
\end{proof}

This result can be extended to general $C^n$ functions with poles of any 
order. In this case the  double integral does not necessarily exist, 
thus we shall use instead the notion of principal value. 

It is not true in general that given a $C^n$ function with poles $g$, the
equation 
$\displaystyle{{\partial \over\partial\bar{z}}f =g}$ 
has a  $C^n$ solution with poles. For instance, $g=1$ does not have 
this property. Any solution of
${\partial \over \partial \bar{z}} f = 1$ is of the form $f(z,\bar z) = 
\bar{z} + p(z)$, where $p$ is a holomorphic function,
and so  $z^m f(1/z,1/\bar z)$ is
not $C^n$ at the origin for any $m \leq n-2,$ i.e. $f$ is not a
$C^n$ function with poles.

\begin{theorem}
Let $g$ be a $C^n$ function with poles. Suppose moreover
that  $\bar{z}^2 \ g$ is also a $C^n$ function
with poles. Then, there exists a
$C^n$ function $f$ with poles such that
$\displaystyle{{\partial \over\partial\bar{z}}f =g}$

\label{existence}
\end{theorem}

\begin{proof}
Let ${\cal F} = \{ z_1,\ldots, z_s\}$ be the poles of $g$ in $\C$,
and suppose $(z - z_i)^{m_i} g(z,\bar z)$ is a $C^n$ function around
each pole $z_i$. Denote $\displaystyle{p(z):= \prod_i (z - z_i)^{m_i}}.$
Then, $p \cdot g$ is a $C^n$ function in the plane.

Since $\bar{z}^2 \ g$ is a $C^n$ function with poles,
there exist a positive number  $R' >0$ and a $C^n$ function $h$
on $B_{2/R'}$ such that 
$h(z,\bar z) := - { z^m g(1/z,1/\bar z)\over \bar{z}^2},$ 
for some $m \leq n-2.$
Consider a $C^\infty$ function $\psi$ with support compact contained
in $B_{2/R'}$ which is identically equal to $1$ on the ball
$B_{1/R'}$ of radius $1/R'$ around the origin. Then,
by Remark \ref{ene} after
Theorem \ref{green}, there exists
a $C^n$ function $H$ on $B_{2/R'}$ such that
$\displaystyle{{\partial \over\partial\bar{z}} H =  \psi \cdot  
h(z,\bar z).}$
Then, the restriction of $H$ to $B_{1/R'}$ is a $C^n$
function satisfying  $\displaystyle{{\partial \over\partial\bar{z}} H =  
h(z,\bar z) }$. 
Set $h_1(z,\bar z) := z^m H(1/z,1/\bar z).$ 
Then, $h_1 \in C^n( |z| > R')$ satisfies condition $2$ in the definition of   
$C^n$ functions with poles. 
Moreover,
$\displaystyle{{\partial \over\partial\bar{z}}h_1 =}$
$\displaystyle{ - {z^m \over
\bar{z}^2} {\partial \over\partial\bar{z}}(H) (1/z,1/\bar z) = }$ $=g$ on
$|z| > R'.$

Let $R > R'$ and, similarly,
take a $C^\infty$ function $\psi$ with support compact contained
in the ball $B_{2R}$ of radius $2 R$ around the origin, such that $\psi \equiv
1 $ on the ball $B_R$ of radius $R.$ Again by Remark \ref{ene} after
Theorem \ref{green}, there exists
a $C^n$ function $H'$ on $B_{2 R}$ such that
$\displaystyle{{\partial \over\partial\bar{z}} H' =  \psi \cdot p \cdot g.}$
Then, $h_2 : = H' / p$ is a $C^n$ function  on $B_R - {\cal F}$ which
satisfies condition $1$ in the definition of $C^n$ functions with poles,
and clearly $\displaystyle{{\partial \over\partial\bar{z}}h_2 =g}$ on
$B_R - {\cal F}.$

Then, $h_2 - h_1$ is a holomorphic function on the annulus
$ R' < |z| < R$, and therefore admits a Laurent expansion
$$ h_2(z,\bar z) - h_1(z,\bar z) = \sum_{- \infty}^{ + \infty} a_k z^k.$$
Denote $\displaystyle{H_2(z):=  -  \sum_{1}^{ + \infty} a_k z^k , \,
H_1(z):=  \sum_{- \infty}^{ 0} a_k z^k.}$  Observe that $H_2$ is a
Taylor series which is convergent for points with absolute value
greater that $R'$ and thus it must be convergent
in the whole ball of radius $R$ around the origin. Similarly,
$H_1$ defines a holomorphic function on $|z| > R'$ (which is also
holomorphic at infinity). Clearly, $ H_1 + h_1 = H_2 + h_2$
on the annulus. Then, we have a well defined global function $f$,
$f = H_2 + h_2$  for $|z| < R$ and 
$f = H_1 + h_1$ for $|z| > R'$,
satisfying $\displaystyle{{\partial \over\partial\bar{z}}f =g}.$ 
Moreover, $f$ satisfies both conditions in Definition \ref{def:pole},
i.e. $f$ is a $C^n$ function with poles. \qed
\end{proof}

\begin{corollary}
Let $g$ be a $C^n$ function  with poles, regular outside the finite
set ${\cal F}$, such that $\bar{z}^2 \ g$ is also a $C^n$ function
with poles.
Then, the function defined in $\C^n - {\cal F}$ by
\begin{equation}
f(z,\bar z)={1\over 2\pi i} {\rm PV}\int_\C g(\zeta,\bar \zeta){\mbox{d}\zeta\wedge 
\mbox{d}\bar{\zeta}\over (\zeta-z)}
\label{Irep}
\end{equation}
is a $C^n$ function with poles and satisfies that 
$\displaystyle{{\partial \over\partial\bar{z}}f =g}$ for all $z \not\in 
{\cal F}$. 
\label{intrep2}
\end{corollary}

\begin{proof}
Since $g(\zeta)$ is a $C^n$ function with poles, ${g(\zeta,\bar \zeta)
\over  \zeta - z}$ is also a $C^n$
function with poles, which ensures the existence of the principal
value. By Theorem \ref{existence}, there exists a $C^n$ function with
poles $F$ such that 
$\displaystyle{{\partial \over \partial\bar{\zeta}}} F = g.$ 
Then, 
$\displaystyle{{\partial \over \partial\bar{\zeta}}} {\left( F(\zeta,\bar \zeta)
\over  \zeta - z \right)} = {g(\zeta,\bar \zeta) \over  \zeta - z }.$
Note that we can deduce from the proof of Theorem \ref{existence}
that the poles of $F$ in the plane are also contained in $\cal F.$
For any $z \in \C - {\cal F},$
$\mbox{Res} \bigg ( {F(\zeta,\bar \zeta)\over (\zeta-z)};z 
\bigg ) = F(z,\bar z). $

By Theorem \ref{cauchy2} we obtain
\begin{equation}
f(z,\bar z)= 
{1\over 2\pi i}{\rm PV}\int_\C  g(\zeta,\bar \zeta)
{\mbox{d}\zeta\wedge \mbox{d}\bar{\zeta}\over  (\zeta-z)} =
\sum_{z_l \in {\cal F} \cup \{\infty\}}\mbox{Res} \bigg ({F(\zeta,\bar \zeta)\over  
(\zeta-z)};z_l\bigg ) + F(z,\bar z).
\label{eq:f-F}
\end{equation}
We deduce from Lemma \ref{lemma2} that each residue in the above
sum is a rational (meromorphic)
function of $z$ (with poles contained in ${\cal F} \cup \{\infty\}$). 
Therefore,  $f$ is a $C^n$ function with poles and
${\partial \over\partial\bar{z}} f = g.$ \qed
\end{proof}

Although the main motivation of this paper is to explicitly show how to 
compute the PV of integrals like (\ref{Irep}), it may be useful to have an 
integral representation of a solution of the differential equation 
${\partial \over\partial\bar{z}} f = g.$ 
Note that this solution is not unique since we  can always add a  
meromorphic function  to the function $f$ defined by (\ref{Irep}).

\section{Integrals on the Sphere}
\label{sec4}
In this section we shall apply the results obtained in section \ref{sec3} 
in order to evaluate integrals on the two sphere whose  integrands  might 
possess singular behavior.  The  main result is stated in Theorem 
\ref{cauchysphere}.
 
Let $(\theta,\phi)$ be the usual coordinates of the sphere, the complex 
stereographic coordinates transformation $z={\rm e}^{i\phi}\cot\left(
\frac{\theta}{2}\right)$ entails that $S^2={\C}\cup \{\infty\}$. 

Now let us assume that  $g$ is  an integrable function on the sphere. Using 
the coordinates $(z,\bar{z})$ defined above,  we  write
\begin{equation}
\int\hspace{-.1in}\int_{S^2} g(\theta,\phi)\sin\theta \mbox{d}\theta\mbox{d}
\phi=2i\int_\C g(z,\bar z){\mbox{d}z\wedge \mbox{d}\bar{z}\over (1+z\bar{z})^2}.
\label{integral}
\end{equation}

The orientation that we use in the complex plane is the right-hand orientation,
i. e. $\mbox{d}z\wedge\mbox{d}\bar{z}=-2i\mbox{d}x \wedge\mbox{d}y$. 
Note that the r.h.s. of (\ref{integral}) is  an improper integral and since 
$g$ is integrable, the principal value of the integral is the value of the 
integral. 

We now consider functions with a finite number of singularities
on the sphere.

\begin{definition}\label{main-def}
Let $\cal F$ be a finite set of points in the sphere.
We say that function $g \in C^n( S^2 -{\cal F})$ 
is a $C^n$ function with poles on $S^2$ if the restriction of $g$
to the complex plane is a $C^n$ function with poles according
to Definition \ref{def:pole}.
\label{def:spherepole}
\end{definition}
 
\begin{remark}
Consider for instance a $C^\infty$ function $g$ on $S^2$ which coincides
with $1/z$ near infinity. Then, $g$ is regular at $\infty$ but
${\rm Res}(g;\infty) \not= 0.$
\end{remark}

\medskip

Given a $C^n$ function $g$ with poles on $S^2$, let 
$D_R$ be a disk centered at the origin, with radius $R$ sufficiently large so 
that the singularities of $g$ remain inside $D_R$. We write 
\begin{eqnarray}
\int\hspace{-.1in}\int_{S^2} g \mbox{d}\mu &:=&2i\int_\C g(z,\bar z){\mbox{d}z\wedge \mbox{d}\bar{z}\over (1+z\bar{z})^2}\nonumber\\
&=&
\lim_{{R\to \infty} \atop {\varepsilon\to 0}}2i\int_{D_R^\varepsilon}  g(z,\bar z){\mbox{d}z\wedge 
\mbox{d}\bar{z}\over (1+z\bar{z})^2}
\label{intC}
\end{eqnarray}
where $D_R^\varepsilon$ is the region obtained deleting  from $D_R$ 
discs of radius $\varepsilon$ around the singularities of $g$.

\medskip

A key observation is that for any $C^n$ function with poles $g$, the function
$G(z, \bar z):= \frac{g}{(1+z\bar z)^2}$ satisfies the additional
hypothesis in Theorem \ref{existence}. In fact, 
given $m$ and a $C^n$ function $h$ around the origin such that
$h(z, \bar z) = z^m g(1/z, 1/\bar z),$ 
${z^m G(1/z,1/\bar z) \over \bar{z}^2} = h(z,\bar z) \ \frac{z^2}
{(1+z\bar z)^2}$ is a $C^n$ function near the origin. 
Therefore, by
Theorem \ref{existence} (or Corollary \ref{intrep2}), there exists a $C^n$
solution with poles $f$ of the differential equation

\begin{equation}
{\partial\over\partial \bar{z}}f=\frac{g}{(1+z\bar z)^2}.
\label{eq:diffeq}
\end{equation}

\begin{remark} 
\label{eq:homog}
We can add to $f$ in (\ref{eq:diffeq}) an arbitrary rational function
$w$ (i.e. an arbitrary solution with pole-singularities of the homogeneous 
differential equation ${\partial\over\partial \bar{z}} w=0).$
\end{remark}

Using   the Green-Stokes formula (\ref{Stokes1}), we obtain
\begin{equation}
\int_{D_R^\varepsilon}{\partial\over\partial \bar{z}}f(z,\bar z)\mbox{d}z\wedge 
\mbox{d}\bar{z}=\sum_l\oint_{\gamma_\varepsilon^l} f(z,\bar z)\mbox{d}z-\oint_{\gamma_R} 
f(z,\bar z)\mbox{d}z,
\label{Stokes3}
\end{equation}
where $\gamma^l_\varepsilon$ and $\gamma_R$ are  circles, the first ones 
centered at the singularities $z=\eta_l$ with radius $\varepsilon$ and the second 
one is the boundary of $D_R$, all with counterclockwise orientation. 

Thus, the integral (\ref{intC}) becomes
\begin{eqnarray}
\int_{S^2}g\mbox{d}\mu&=&
\lim_{{R\to \infty} \atop {\varepsilon\to 0}}2i\int_{D_R^\varepsilon}  
{\partial\over\partial \bar{z}}f(z,\bar z)\mbox{d}z\wedge \mbox{d}\bar{z}\nonumber\\
&=&2i\sum_l\lim_{\varepsilon\to 0}\oint_{\gamma_\varepsilon^l} f(z,\bar z)\mbox{d}z-2i
\lim_{R\to \infty}\oint_{\gamma_R} 
f(z,\bar z)\mbox{d}z.
\label{Stokes4}
\end{eqnarray}

As a consequence of Theorem \ref{cauchy2} we get the following result.

\begin{theorem}\label{main-theorem}
Assume that  $g$  is a  $C^n$ function with poles on $S^2$ 
and let $f$ be a $C^n$ function with poles  satisfying (\ref{eq:diffeq}). 
Then the {\it principal value} 
$$
{\rm PV}\int_{S^2}g\mbox{d}\mu=2i\; {\rm PV}  
\int_{\C} \frac{g(z,\bar z)}{(1+z\bar z)^2}\mbox{d}z\wedge \mbox{d}\bar{z}
$$
exists and  
$$
{\rm PV}\int_{S^2}g\mbox{d}\mu= 
-4\pi \bigg (\sum_l{\rm Res}(f;\eta_l
)+{\rm Res}(f;\infty)
\bigg).
$$
where $\eta_l$ are the poles of $f$.
\label{cauchysphere}
\end{theorem}

The proof follows at once from Theorem \ref{cauchy2}. Besides if
$g$ is  integrable,  {\it the principal value} is the value of the integral. 

Note that as we said in Remark \ref{eq:homog} the function $f$ is not unique, 
but the result given above is independent of the choice of the solution of 
(\ref{eq:diffeq}) since the sum of the residues of a meromorphic function on 
$S^2$ is zero.

The same ideas as in Corollary \ref{intrep2} can be used 
to present an explicit solution with poles of the
differential equation (\ref{eq:diffeq}). 

\begin{corollary}\label{main-cor}
Let $g$ be a $C^n$ function with poles on $S^2$, regular
outside ${\cal F} = \{z_1,\ldots, z_s\}.$
Then the function defined in $S^2-{\cal F}$
$$
f(z,\bar z)={1\over 2\pi i} {\rm PV}\int_\C \frac{g(\zeta,\bar \zeta)}
{(1+\zeta\bar\zeta)^2}{\mbox{d}\zeta\wedge 
\mbox{d}\bar{\zeta}\over (\zeta-z)}.
$$
is a $C^{n}$ function with poles on $S^2$ and satisfies that 
$\displaystyle{{\partial \over\partial\bar{z}}f =
\frac{g}{(1+z\bar z)^2}}$ for all \linebreak
$z \in \C, \, z\neq z_j, \, j=1,\ldots,s.$
\label{intrep3}
\end{corollary}

Finally, we exemplify the result given in Theorem \ref{cauchysphere}
for $g=1$ and different  choices of $f$. Consider the integral
$$
\int\hspace{-.1in}\int_{S^2} \sin\theta\mbox{d}\theta\mbox{d}\phi= 2i 
\int_{\bf C}{\mbox{d}z\wedge \mbox{d}\bar{z}\over (1+z\bar{z})^2},
$$
and the solutions of (\ref{eq:diffeq}) $f_0(z,\bar z)=-{1\over z(1+z\bar z)}$ and  
$f_1(z,\bar z)=\frac{\bar z}{1+z\bar z}$, 
then
$$
{\rm Res}(f_0;0)=-1\qquad \mbox{and}\qquad {\rm Res}(f_0;\infty)=0,
$$
thus the contribution to the integral comes from the pole of $f_0$ at zero. 
Whereas the contribution to the integral using $f_1$  comes from the  residue 
at infinity. Note the the function $f_0$ has a pole at $z=0$ whereas $f_1$ 
is $C^\infty({\bf C})$, moreover  $f_1$  is the integral representation given
by Theorem \ref{intrep}.

\subsection*{Acknowledgments}
We are  indebted
to  Carlos N. Kozameh for many enlightening conversations
and suggestions, and  to Simonetta Frittelli  for  many useful comments.
We would also like to thank an anonymous referee whose observations
led to a considerably improved version of this paper.

\end{document}